\documentclass[prc,aps,amsfonts,amsmath,superscriptaddress,floatfix,twocolumn]{revtex4}
\usepackage{graphicx,float} 
\usepackage{color}
\usepackage{mwe}
\usepackage{braket}
\usepackage[caption=false]{subfig}
\usepackage{hyperref}
\usepackage{ulem}

\begin{document}

\title{Low-energy cluster vibrations in N = Z nuclei}

\author{F. Mercier}
\affiliation{IJCLab, Universit\'e Paris-Saclay, IN2P3-CNRS, F-91406 Orsay Cedex, France}

\author{A. Bjel\v{c}i\'{c}}
\affiliation{Department of Physics, Faculty of Science, University of Zagreb, Bijeni\v{c}ka c. 32, 10000 Zagreb, Croatia}

\author{T. Nik\v{s}i\'{c}}
\affiliation{Department of Physics, Faculty of Science, University of Zagreb, Bijeni\v{c}ka c. 32, 10000 Zagreb, Croatia}

\author{J.-P. Ebran}
\affiliation{CEA,DAM,DIF, F-91297 Arpajon, France}
\affiliation{Universit\'e Paris-Saclay, CEA, Laboratoire Mati\`ere en Conditions Extr\^emes, 91680, Bruy\`eres-le-Ch\^atel, France}

\author{E. Khan}
\affiliation{IJCLab, Universit\'e Paris-Saclay, IN2P3-CNRS, F-91406 Orsay Cedex, France}

\author{D. Vretenar}
\affiliation{Department of Physics, Faculty of Science, University of Zagreb, Bijeni\v{c}ka c. 32, 10000 Zagreb, Croatia}


\begin{abstract} 
Significant transition strength in light $\alpha$-conjugate nuclei at low energy, typically below 10 MeV, has been observed in many experiments. In this work the isoscalar low-energy response of N=Z nuclei is explored using the Finite Amplitude Method (FAM) based on the microscopic framework of nuclear energy density functionals. Depending on the multipolarity of the excitation and the equilibrium deformation of a particular isotope, the low-energy strength functions display prominent peaks that can be attributed to vibration of cluster structures: $\alpha$+$^{12}$C+$\alpha$ and $\alpha$+$^{16}$O in $^{20}$Ne, $^{12}$C+$^{12}$C in $^{24}$Mg, 4$\alpha$+$^{12}$C in $^{28}$Si, etc. Such cluster excitations are favored in light nuclei with large deformation.
\end{abstract}
 

\maketitle
\section{Introduction}
%
A number of experiments have observed a significant increase of the E0 strength at excitation energies below the giant monopole resonance in relatively light nuclei~\cite{you97,you98,lui01,you02,yil06,gup15}. Theoretical studies using, e.g., the cluster model~\cite{tom78,ueg79,kam81,des87,suz89,che07,ito11,del12,kan19}, or the Antisymmetrized Molecular Dynamics (generally combined with Generator Coordinate Method (GCM))~\cite{kim04,chi15,chi16,kan16,kan20}, consistently interpret these observations as excitations of cluster structures. Cluster excitations can also occur with higher multipoles~\cite{you99,lui01,you09,che09,itoh11}. For instance, a low-energy E1 excitation has been associated with a reflection-asymmetric vibration of an $\alpha$ cluster against the $^{16}$O core in $^{20}$Ne ~\cite{chi16,kan19}, with a strength that is enhanced in comparison to similar excitations contributing in the E0 and E2 response.

Valuable information about the structure of a nucleus can be obtained by analyzing how the system responds to an external perturbation with a given multipolarity (see, for instance, \cite{nak19,per14}). A useful theoretical framework for such studies is provided by the Random-Phase Approximation (RPA), and the Quasiparticle-RPA (QRPA) which extends the former to superfluid systems. (Q)RPA calculations on top of reference mean-field states computed using Energy Density Functionals (EDFs), have demonstrated the capacity to describe excitation modes ranging from tens of keV to tens of MeV \cite{kha11,per07}. The method has also been extended to charge-exchange modes \cite{fra05,mar14,lia12}. There are many ways to derive the QRPA equations, e.g. by linearizing the Hartree-Fock-Bogoliubov (HFB) equations and then solving an eigenvalue problem ~\cite{rin80}. A major issue in QRPA calculations are the dimensions of the matrix system which can become very large, especially when the HFB reference state is allowed to spontaneously break the symmetries of the nuclear Hamiltonian.

Several methods have been developed to circumvent these numerical difficulties~\cite{yos08,pen09,ter10,los10}, here in particular we focus on the Finite Amplitude Method (FAM)~\cite{nak07}. It is also based on the linearization of the Hartree-Fock (HF) equations but avoids the solution of a matrix eigenvalue problem. The FAM has been extended to superfluid systems  (QFAM)~\cite{avo11} for Skyrme interactions and relativistic functionals \cite{lia13,nik13}. The Skyrme-based FAM has been applied to photoabsorption cross sections~\cite{ina09}, higher multipole excitation modes~\cite{kor15}, giant dipole resonances in heavy nuclei~\cite{tom16}, and $\beta ^{-}$ decay studies~\cite{mus16}.

The present study is based on the relativistic QFAM ~\cite{nik13}. Relativistic EDFs have successfully been used to describe both liquid- and cluster-like nuclear properties \cite{ebr12,ebr14,ebr14,ebr14b,ebr18}, starting from nucleonic degrees of freedom. Recently the multi-reference implementation of the GCM based on relativistic EDFs has been employed in the analysis of spectroscopic properties (energies of excited states, elastic and inelastic form factors) of nuclei with cluster structures \cite{mar18,mar19}. A QFAM approach based on relativistic EDFs is hence expected to provide an alternative consistent and microscopic description of cluster vibrations in nuclei. 

In this work we perform a systematic calculation of isoscalar multipole ($\lambda=0,1,2,3$) strength in $\alpha$-conjugate nuclei from $^{12}$C to $^{56}$Ni, and analyze the low-energy structure of the strength functions. The calculations are based on the DD-PC1 parametrization~\cite{Niksic.2008_PRC-78} and involve an expansion of the equations of motion in an axially-deformed harmonic oscillator basis. The first nucleus to be analyzed is $^{20}$Ne whose large equilibrium deformation favors  clusterization, and hence cluster vibration modes are expected to occur at low energy~\cite{kim04}. We will show that the lowest modes correspond to reflection-symmetric $2 \alpha + ^{12}$C and reflection-asymmetric $\alpha + ^{16}$O configurations oscillating around the axially-symmetric deformed equilibrium. The study of $^{20}$Ne is extended to other $\alpha$-conjugate nuclei, and the evolution of the strength function is analyzed when the quadrupole moment of the mean-field reference state is varied from oblate to prolate deformations.
 
The QFAM formalism is briefly introduced in Sec. II. Section III explores the multipole ($\lambda=0,1,2,3$) response of $^{20}$Ne, as well as the role played by quadrupole deformation in the appearance of cluster vibration modes. 
In Sec. IV we extend the study of isoscalar monopole vibrations to three other $\alpha$-conjugate nuclei that display pronounced cluster vibrations: $^{24}$Mg, $^{28}$Si and $^{32}$S.
Section V contains a brief summary and conclusions.
%
\section{Theoretical framework}
%
Our implementation of the QFAM follows closely the one described in Refs.~\cite{Stoitsov.2011_PRC-84,Kortelainen.2015_PRC-92}.
The QFAM equations read:
\begin{align}
\label{Eq:QFAM-20}
(E_\mu + E_\nu - \omega) X_{\mu \nu}(\omega) + \delta H_{\mu \nu}^{20}(\omega)  &= -F_{\mu \nu}^{20},\\
\label{Eq:QFAM-02}
(E_\mu + E_\nu + \omega) Y_{\mu \nu}(\omega) + \delta H_{\mu \nu}^{02}(\omega)  &= -F_{\mu \nu}^{02},
\end{align}
where the matrices $F^{20}$ and $F^{02}$ are calculated from the external harmonic perturbation field:
\begin{equation}
F(t) = \eta \left(F(\omega) e^{-i\omega t} + F^\dagger(\omega) e^{+i\omega t}   \right),
\end{equation} 
characterized by the small real parameter $\eta$.
$X_{\mu \nu}(\omega)$ and 
$Y_{\mu \nu}(\omega)$ denote the QFAM amplitudes at given excitation energy $\omega$, while 
$\delta H_{\mu \nu}^{20}(\omega)$ and $\delta H_{\mu \nu}^{02}(\omega)$ describe the response of the atomic nucleus to the
external perturbation.
The time-dependent density matrix and pairing tensor read:
\begin{align}
\rho(t) &= V^* V^T + \eta \left(\delta \rho(\omega) e^{-i\omega t} + \delta \rho^\dagger(\omega) e^{+i\omega t}\right),\\
\kappa(t) &= V^*U^T + \eta \left(  \delta \kappa^{(+)}(\omega) e^{-i\omega t} + \delta \kappa^{(-)}(\omega) e^{+i\omega t}  \right),
\end{align}
where
\begin{align}
\delta \rho(\omega) &= U X(\omega) V^T + V^* Y^T(\omega) U^\dagger,\\
\delta \kappa^{(+)}(\omega) &= U X(\omega) U^T + V^* Y^T(\omega) V^\dagger,\\
\delta \kappa^{(-)}(\omega) &= V^* X^\dagger(\omega) V^\dagger + U Y^*(\omega) U^T.
\end{align}
The transition strength at each particular energy is calculated from the expression:
\begin{equation}
S(f,\omega) = -\frac{1}{\pi} \textnormal{Im} \textnormal{Tr} \left[ f^\dagger \delta \rho(\omega) \right],
\end{equation}
where $\delta \rho(\omega)$ denotes the induced density matrix, and $f_{kl}$ are the matrix elements
of the operator $F(\omega)$ in configuration space. 

To prevent that the QFAM solutions diverge in the vicinity of a QRPA state, a
small imaginary part is added to the energy $\omega \to \omega + i \gamma$. This corresponds to  folding the QRPA strength function with a Lorentzian of width $\Gamma = 2\gamma$~\cite{Avogadro.2011_PRC-84}. 
The electric isoscalar multipole operator is defined as 
\begin{equation}
\displaystyle f_{JK}^{IS} = \sum_{i=1}^A{f_{JK}(\mathbf{r}_i)},
\end{equation}
with $f_{JK}(\mathbf{r}) = r^J Y_{JK}(\theta,\phi)$. For the monopole mode the operator reads 
$f_{00}(\mathbf{r}) = r^2$, while for the isoscalar dipole excitation $f_{1K}(\mathbf{r}) = r^3 Y_{1K}(\theta,\phi)$ . 
Since for an even-even axially symmetric nucleus the operators $f_{JK}$ and $f_{J-K}$ produce 
identical strength functions, in the code we employ the operator $f^{(+)}_{JK} = \left( f_{JK} + (-1)^Kf_{J-K} \right)/\sqrt{2+2\delta_{K0}}$
and assume $K\ge 0$. 

The DIRQFAM solver is based on the relativistic Hartree-Bogoliubov model 
with the particle-hole channel parametrized by the DD-PC1 energy density functional~\cite{Niksic.2008_PRC-78} , while the particle-particle channel is
determined by a pairing force separable in momentum space~\cite{dug04,Tian.2009_PhysLettB-676}: 
$\displaystyle \langle k | V^{^1S_0}|k^\prime\rangle = -G p(k)p(k^\prime)$. By assuming a simple Gaussian ansatz
$p(k) = e^{-a^2k^2}$, the two parameters G and a
were adjusted to reproduce the density dependence of the pairing gap
at the Fermi surface in nuclear matter obtained by the Gogny D1S interaction~\cite{d1s}. 
The current implementation of the DIRQFAM solver employs an expansion of the Dirac spinors in terms of eigenfunctions of an axially symmetric
harmonic oscillator potential. Further details on the QFAM solver DIRQFAM can be found in Ref.~\cite{Bjelcic.2020_CPC}.
%
\section{Isoscalar vibrations in ${^{20}\rm{Ne}}$}
\begin{figure}[!htb]
\centering
\includegraphics[width=0.95\linewidth]{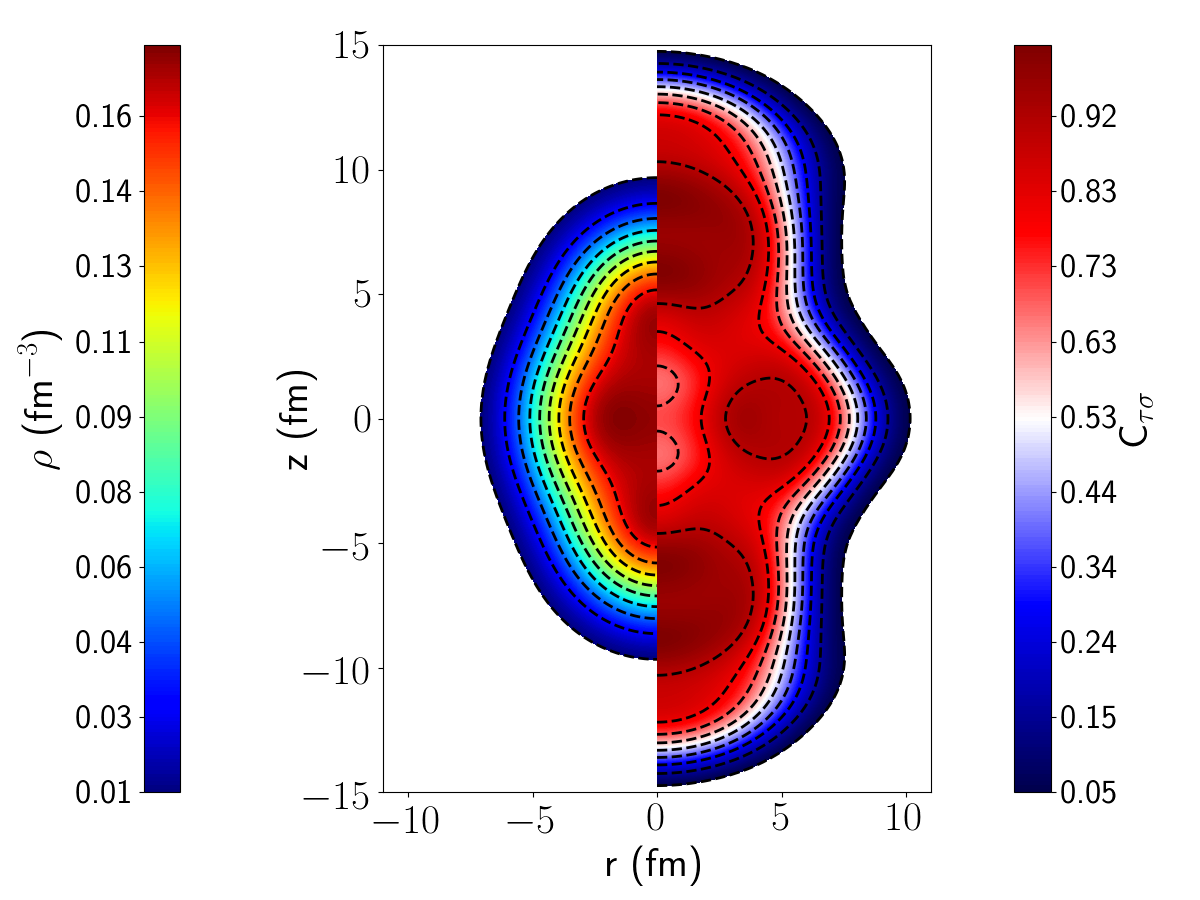}
\caption{(Color online) The self-consistent equilibrium density of $^{20}$Ne (left panel), and localization function $C_{\tau\sigma}$ (right panel) obtained using the RHB model with the DD-PC1 energy density functional.}
\label{fig:Density_20Ne_GS}
\end{figure}

We begin our analysis with the isotope $^{20}$Ne. The left panel of Fig.~\ref{fig:Density_20Ne_GS} displays the prolate deformed ($\beta_2 \approx 0.5$) ground-state intrinsic density of $^{20}$Ne
obtained obtained with the DD-PC1 parametrization. The density exhibits cluster structures at the outer ends of the symmetry axis with density peaks 
$ \simeq 0.2$ fm$^{-3}$, and an oblate deformed core, reminiscent of a quasimolecular $\alpha$-$^{12}$C-$\alpha$ structure. 
The spatial localization and cluster formation in atomic nuclei can also be quantified by using the localization function $C_{\tau\sigma}(\mathbf{r})$, defined in
Ref.~\cite{Reinhard.2011_PRC-83} for the nuclear case. A value of the localization measure close to 0.5 signals that nucleons are delocalized, while a value close to one corresponds to a localized alpha-like structure at point $\vec{r}$ in an even-even $N=Z$ nucleus.
The localization function for $^{20}$Ne is plotted in the right panel of Fig.~\ref{fig:Density_20Ne_GS}, and consistently confirms the alpha-like nature of the localized structures appearing in the density. 

\begin{figure}[!hbt]
      \centering
      \includegraphics[width=0.75\linewidth]{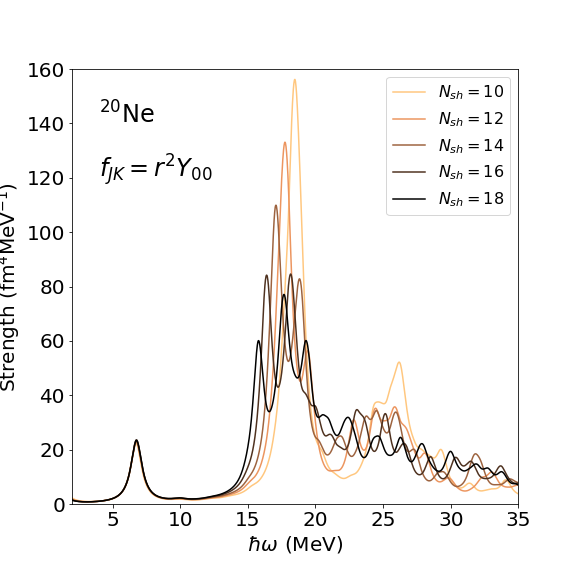}   
    \caption{(Color online) Evolution of the monopole strength function in $^{20}$Ne with the size of the harmonic oscillator basis.}
    \label{fig:evolshell}
  \end{figure}

The isoscalar strength function of the monopole operator $\displaystyle \sum_{i=1}^A{r_i^2}$ for $^{20}$Ne is analyzed using the QFAM. 
The calculation has been performed in the harmonic oscillator basis with $N^{(f)}_{sh}=10,12,14, 16$ and $18$ major oscillator shells for the upper component, and 
$N^{(g)}_{max}=N^{(f)}_{sh}+1$ for the lower component of the Dirac spinor (see Ref.~\cite{Gambhir.91_Annals-90}). In the following discussion the number of shells $N_{sh}$ corresponds to the number of major harmonic oscillator shells used in the expansion of the upper component of the Dirac spinor, i.e., $N_{sh} \equiv N^{(f)}_{sh}$. 
In Fig.~\ref{fig:evolshell} we compare the strength functions of the isoscalar monopole operator for $^{20}$Ne, calculated with 
$N_{sh}=$10, 12, 14, 16 and 18. The low-energy part of the strength
function is fully converged even for relatively small values of the $N_{sh}$. However, for higher energies, the strength function displays a pronounced
dependence on the dimension of the harmonic oscillator basis, essentially because these excitations involve states in the continuum. 
Therefore, the high-energy part of the strength function is strongly affected by the details of single-particle configurations. We note, however, that the centroids of the strength distribution in the high energy region are much less sensitive to the basis dimension, as shown in Tab.~\ref{Tab:centroids}. 
 Since this study is focused on the properties of low-lying states, all subsequent calculations are performed by expanding the large component of the Dirac spinors in $N^{(f)}_{sh}=$14 major oscillator shells.
 
\begin{table}[!hbt]
\begin{tabular}{ccc}
$N_{sh}$ & $\bar{E}_{low}$ (MeV)& $\bar{E}_{high}$ (MeV) \\ \hline 
10 & 18.4 & 27.0 \\
12 & 18.1 & 27.0 \\
14 & 18.1 & 27.3 \\
16 & 18.0 & 27.6 \\
18 & 18.1 & 28.0 
\end{tabular}
\caption{\label{Tab:centroids}Centroids of the monopole strength function (see Fig.~\ref{fig:evolshell}) defined as the ratio of moments $m_1/m_0$. 
The moments
of the strength function are $m_k=\int{E^kS(E)dE}$. The $\bar{E}_{low}$ and $\bar{E}_{higs}$ centroids are calculated in the energy intervals 
10 MeV $\le E \le$ 22.5 MeV and   22.5 MeV $< E \le$ 35 MeV, respectively.}
\end{table}

\begin{figure}[!htb]
\centering
\includegraphics[width=1\linewidth]{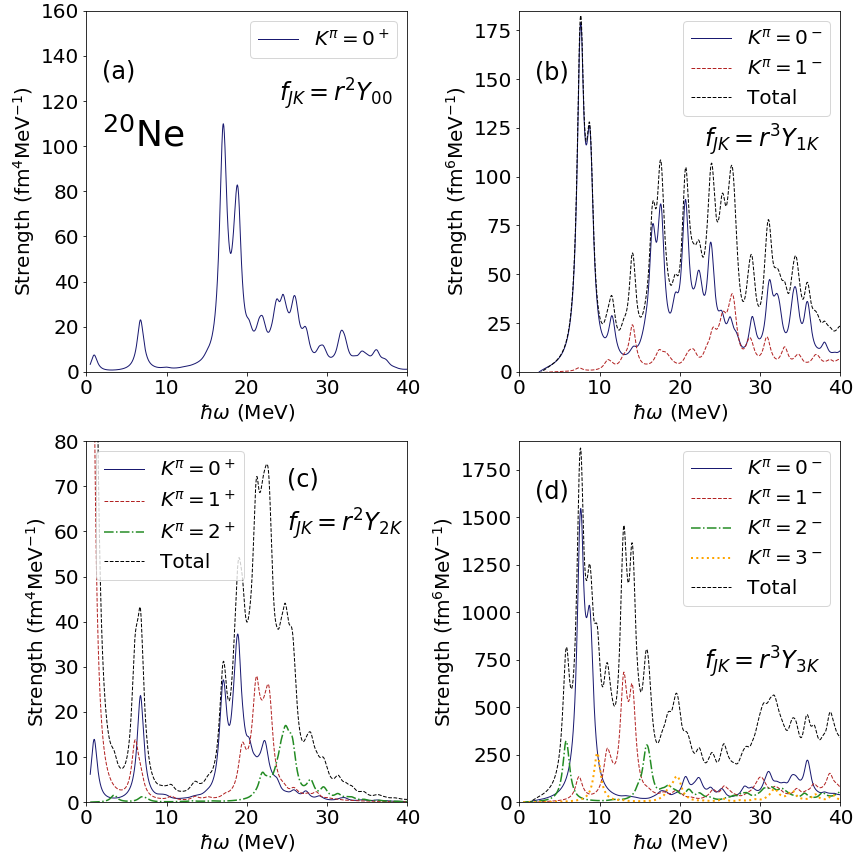}
\caption{(Color online) $^{20}$Ne strength distribution functions for the QFAM response to the isoscalar monopole (panel (a)), 
isoscalar dipole (panel (b)), isoscalar quadrupole (panel (c)) and isoscalar octupole (panel (d)) operator. For $J>0$ multipoles,
the corresponding projections $K=0$ (solid blue), $K=1$ (dashed red), $K=2$ (dot-dashed green) and
$K=3$ (dotted orange) are plotted separately. The thin dashed curves denote the total strength.}
\label{fig:Ne20_strength_all}
\end{figure}

Fig.~\ref{fig:Ne20_strength_all} displays the strength functions for the QFAM response to the isoscalar monopole (panel (a)), 
isoscalar dipole (panel (b)), isoscalar quadrupole (panel (c)) and isoscalar octupole (panel (d)) operator. In addition to the $K = 0$ components, 
for the multipoles $\lambda = 1, 2, 3$ we also plot the contributions of the higher-$K$ projections separately, as well as the total strenghts. 
For the quadrupole $K=1^{+}$ strength distribution one notices the appearance of the spurious state related to the
breaking of rotational symmetry, and also the ordering of the $K=0^{+}$, $K=1^{+}$, and $K=2^{+}$ peaks in the high energy region above 15 MeV 
is consistent with the prolate deformed ground state of $^{20}$Ne. 
Although all strength distributions exhibit pronounced fragmentation in the $E \ge 10$ MeV region, a sizeable portion of strength is  
located at $E \approx 7$ MeV. We have verified that for all multipoles these low-energy peaks are stable with respect to the number of oscillator shells used in the basis expansion.

\begin{figure}[!htb]
\centering
\includegraphics[width=1\linewidth]{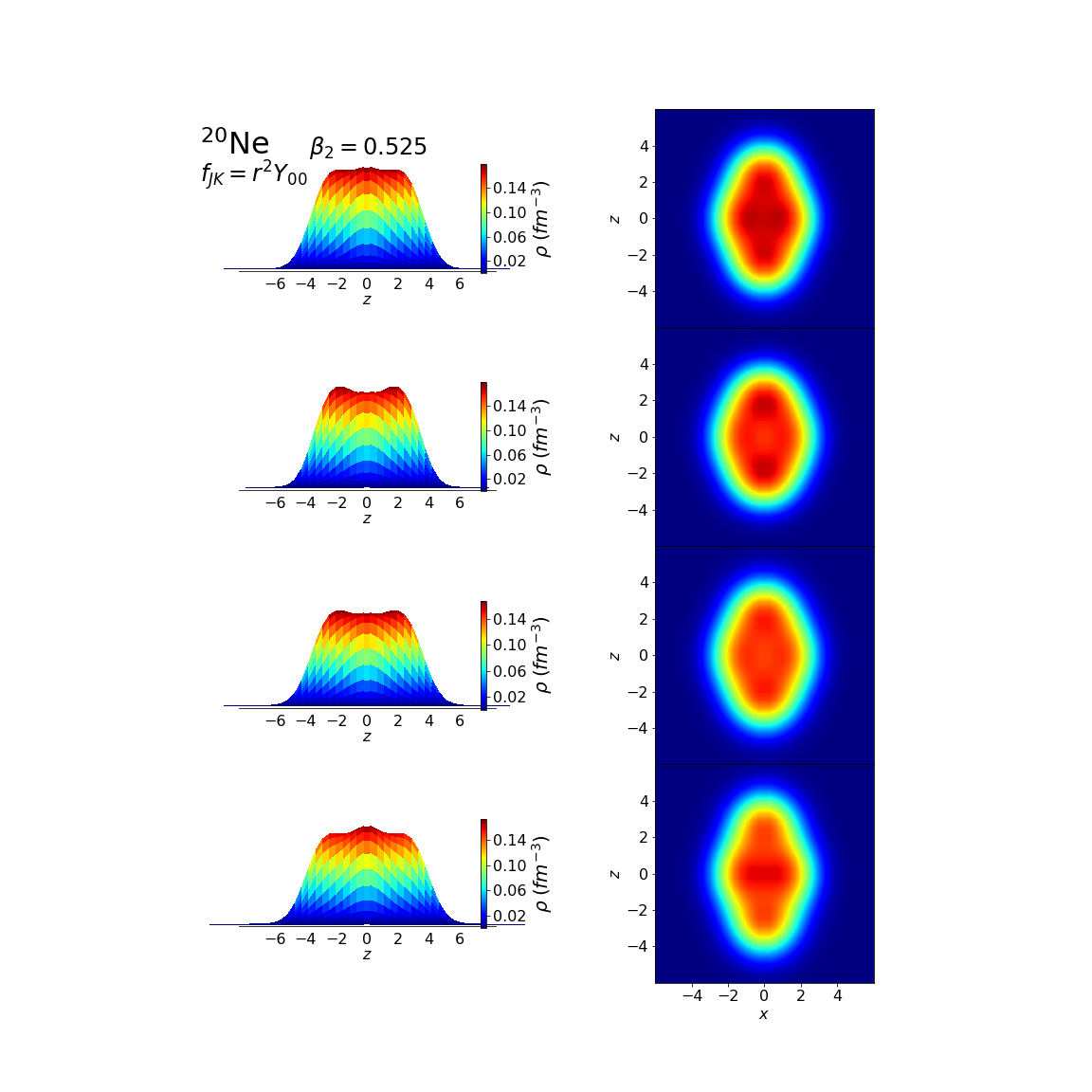}
\caption{(Color online) Snapshots of the $^{20}$Ne density oscillations at energy $\hbar \omega=6.75$ MeV induced by monopole perturbation. Time increases from top to bottom and a full period is shown.
}
\label{fig:Densities_00}
\end{figure}

\begin{figure}[!htb]
\centering
\includegraphics[width=1\linewidth]{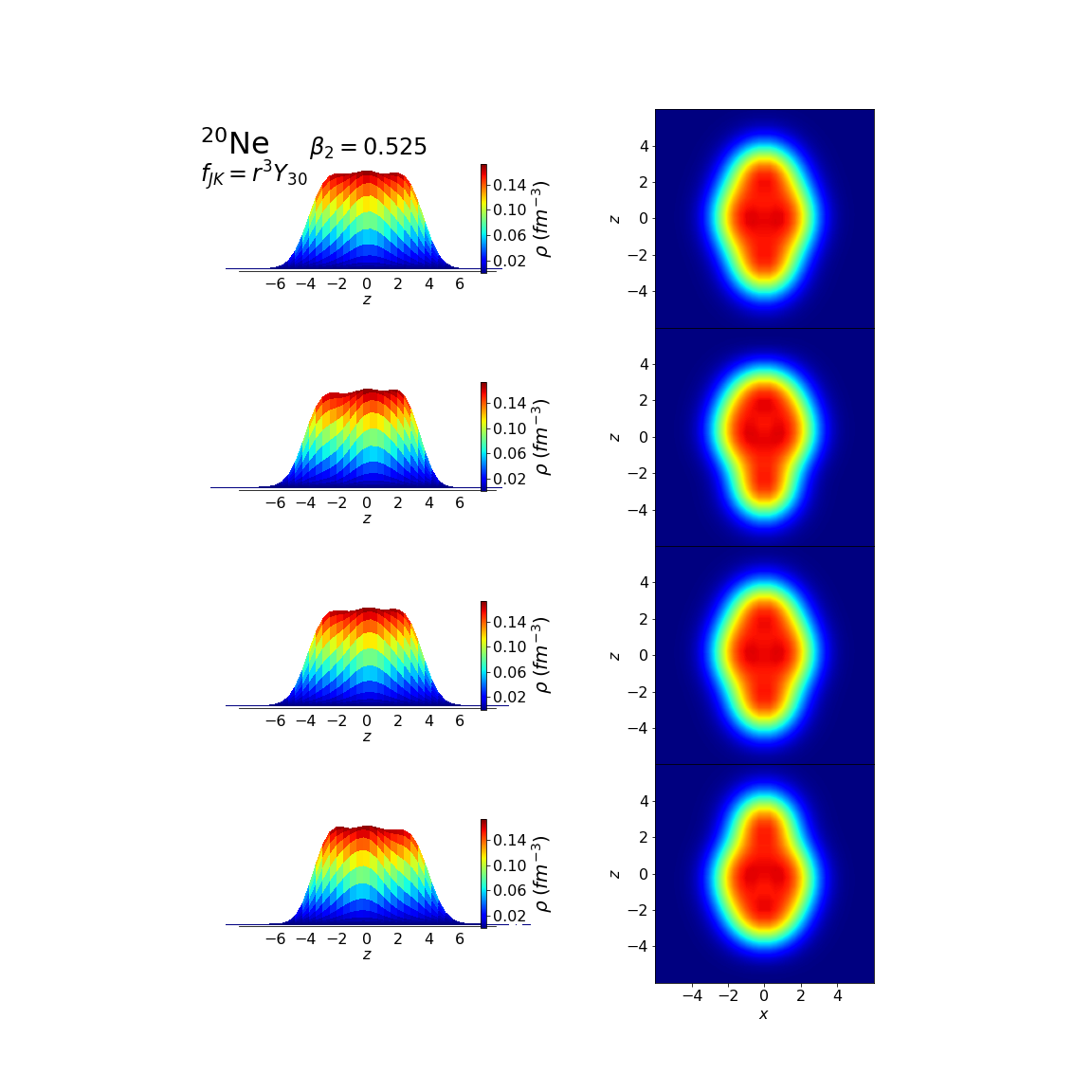}
\caption{(Color online) Snapshots of the $^{20}$Ne density oscillations at energy $\hbar \omega=7.65$ MeV induced by octupole perturbation ($K=0$ component). Time increases from top to bottom and a full period is shown.
}
\label{fig:Densities_30}
\end{figure}

The nature of the low-energy excitations can be analyzed by considering the corresponding transitions densities.
The time-dependent density reads
\begin{equation}
\rho(\mathbf{r},t) = \rho_{gs}(r_\perp,z) + 2\eta \textnormal{Re}\left[e^{-i\omega t} \delta \rho(\omega,r_\perp,z) \right]\cos{(K\phi)},
\label{eq:rhot}
\end{equation}
where $\rho_{gs}(r_\perp,z)$ denotes the ground-state density and $\delta \rho(\omega, r_\perp,z)$ is the transition density at a given excitation energy $\omega$.
We note that for the $K=0$ modes the time-dependent densities are axially symmetric $\delta \rho(\mathbf{r}) = \delta \rho(r_\perp,z)$, hence it is sufficient to
study their behaviour in the $xz$ plane. Figures~\ref{fig:Densities_00} and~\ref{fig:Densities_30} display the snapshots of the
time-dependent density in the $xz$ plane for the low-energy modes induced by monopole and octupole ($K=0$ component) perturbations. Time increases from the top to the bottom, with the time step $\Delta t = 2\pi/4$. The parameter $\eta$ defined by  
Eq.~(\ref{eq:rhot}) equals 0.05 for the monopole and 0.005 for the octupole perturbation, respectively. The large value of the intrinsic equilibrium deformation of $^{20}$Ne leads to cluster formation already in its ground state, and one finds that clusters oscillate against the core for both modes shown in Figs.~\ref{fig:Densities_00} and~\ref{fig:Densities_30}. Furthermore, two different types of vibrations are observed: i) the two $\alpha$ clusters oscillate against the $^{12}$C core for the $J=0$ reflection-symmetric mode, ii)
an oscillation of the $\alpha$ cluster against the $^{16}$O core for the $J=3$ reflection-asymmetric mode.

%

\begin{figure}[!htb]
\centering
\includegraphics[width=0.75\linewidth]{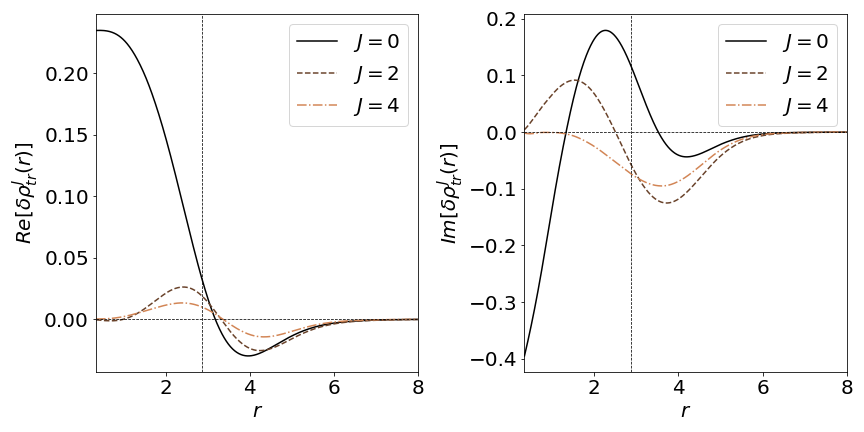}
\caption{(Color online) Radial parts of the angular-momentum projected transition densities that correspond to the 
low-energy peak of the isocalar monopole response of $^{20}$Ne. The real and imaginary parts of the transition density 
are shown in the left and right panels, respectively. The ground state $rms$ radius is indicated by the vertical dashed line.}
\label{fig:td_radial_mon}
\end{figure}

\begin{figure}[!htb]
\centering
\includegraphics[width=0.75\linewidth]{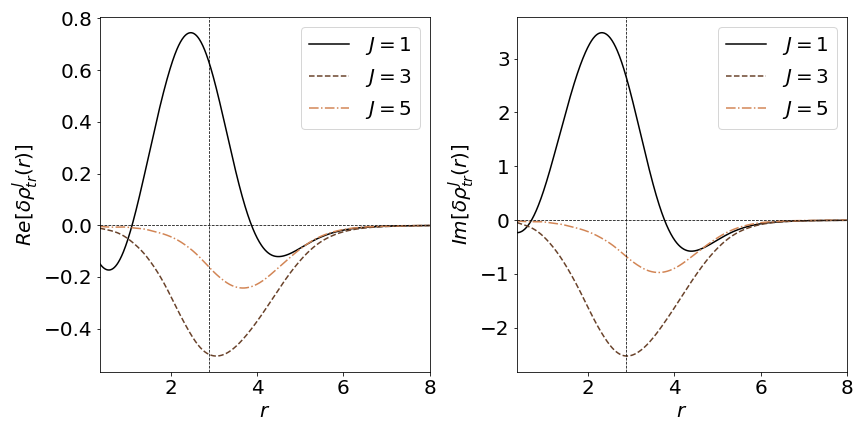}
\caption{(Color online) Same as in the caption to Fig.~\ref{fig:td_radial_mon} but for the 
isocalar octupole response ($K=0$ component).}
\label{fig:td_radial_oct}
\end{figure}

The two-dimensional intrinsic transition densities $\delta \rho_{tr}(\mathbf{r})$ can be projected onto good angular momentum to yield the transition densities in the 
laboratory frame of reference. For a particular value of the angular momentum $J\ge K$, the two dimensional projected transition density can be approximated using its radial part by
\begin{equation}
\delta \rho_{tr}^J(\mathbf{r}) = \delta \rho_{tr}^{J}(r) Y_{JK}(\Omega), 
\end{equation}
with the radial part defined as
\begin{equation}
\delta \rho_{tr}^J(r)  = \int{d\Omega \delta \rho_{tr}(r_\perp,z)Y^*_{JK}(\Omega)} .
\end{equation}
Fig.~\ref{fig:td_radial_mon} compares the radial parts of the angular-momentum-projected transition densities 
$\delta\rho_{tr}^{J=0}(r)$, $\delta\rho_{tr}^{J=2}(r)$ and $\delta\rho_{tr}^{J=4}(r)$ that correspond the the low-energy peak of the isocalar monopole response
in $^{20}$Ne. The real and imaginary parts of the transition density are displayed in the left and right panels, respectively. For the real parts we note the characteristic node of the transition density close to the position of the $rms$ radius. The radial parts of the angular-momentum-projected transition densities 
$\delta\rho_{tr}^{J=1}(r)$, $\delta\rho_{tr}^{J=3}(r)$ and $\delta\rho_{tr}^{J=5}(r)$ that correspond the the low-energy peak of the isocalar octupole response are shown in Fig.~\ref{fig:td_radial_oct}. In contrast to the volume monopole mode, the isoscalar octupole transition densities exhibit the predominantly surface nature of the octupole mode. 

\begin{figure}[!htb]
\centering
\includegraphics[width=1\linewidth]{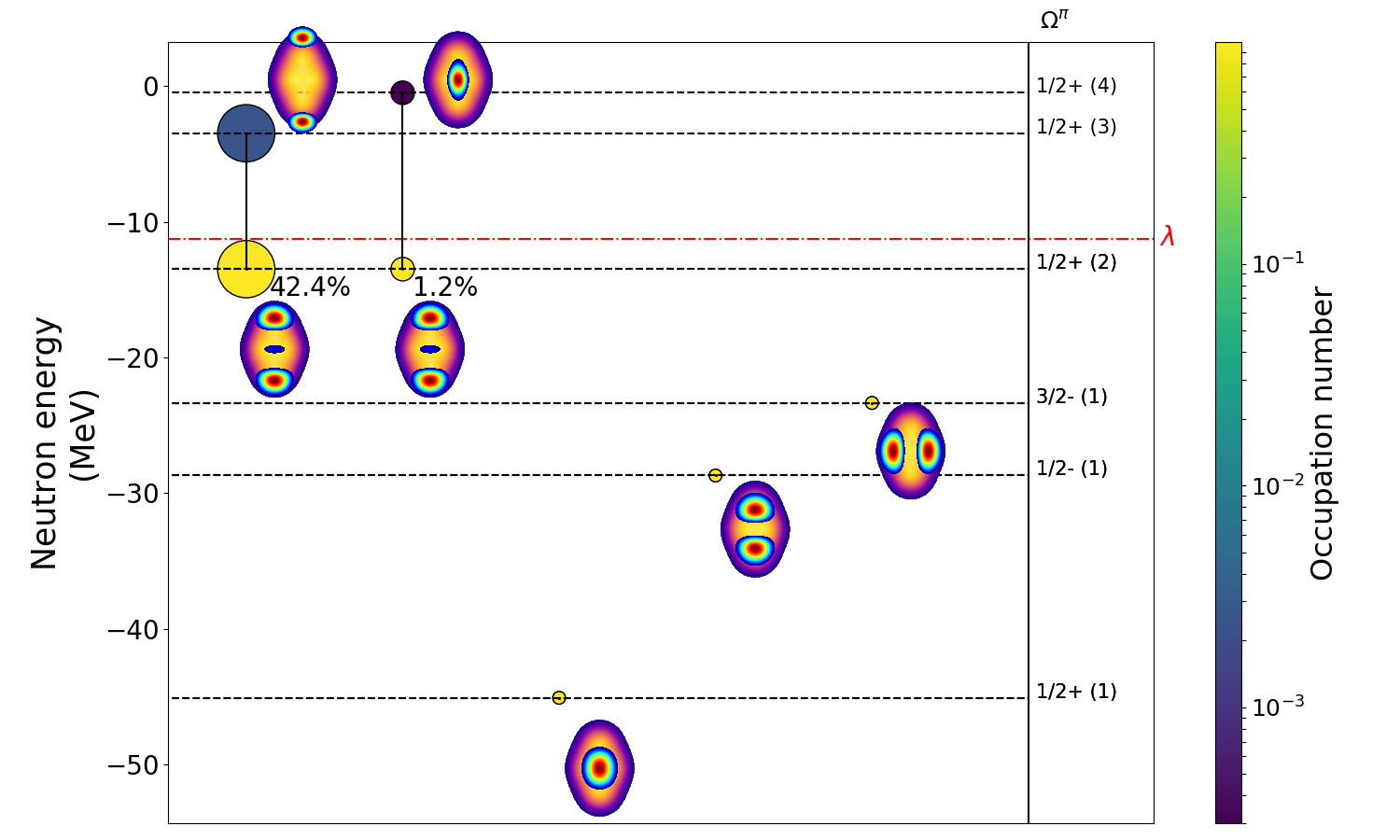}
\caption{(Color online) Schematic illustration of the most important neutron 2-qp contributions to the isoscalar monopole excitation at      $\hbar \omega = 6.7$ MeV in $^{20}$Ne.  The area and the number below represent the fraction of the total $|X|^2 - |Y|^2$ (see Eq.\ref{eq:XY}) for this particular excitation. The $\Omega \pi$   quantum numbers are listed on the right of the figure. The associated partial densities are also plotted for each of the configurations as well as the total density in the background. The Fermi level is shown as a red dash-dotted line.
}
\label{fig:excitation_config}
\end{figure}

It is instructive to decompose the excitation modes in terms of 2-quasiparticle (2-qp) contributions~\cite{Deloncle.17_EPJA-53}. 
This can be achieved by using the contour integration procedure introduced in 
Ref.~\cite{Hinohara.2013_PRC-87}. The individual QRPA amplitudes corresponding to the excitation mode $i$ are calculated as
\begin{align}
X_{\mu\nu}^i &= e^{-i\theta} | \langle i | \hat{F}| 0 \rangle |^{-1} \frac{1}{2\pi i} \oint_{C_i}{X_{\mu\nu}(\omega_\gamma)d\omega_\gamma},\\
Y_{\mu\nu}^i &= e^{-i\theta} | \langle i | \hat{F}| 0 \rangle |^{-1} \frac{1}{2\pi i} \oint_{C_i}{Y_{\mu\nu}(\omega_\gamma)d\omega_\gamma},
\end{align}
where $X_{\mu\nu}(\omega_\gamma)$ and $Y_{\mu\nu}(\omega_\gamma)$ denote the QFAM amplitudes for the complex frequency $\omega_\gamma = \omega+i\gamma$, 
and $C_i$ is the contour in the complex energy plane that encloses the first-order pole on the real axis at $\omega_\gamma = \Omega_i$. 
We note that the common phase $e^{i\theta}$ remains arbitrary. The individual 2-qp contributions to some particular excitation mode $i$ can be quantified by the following quantity:
\begin{equation}
\xi^i_{2qp}=\left|X^i_{2qp}\right|^{2}-\left|Y^i_{2qp}\right|^{2}.
\label{eq:XY}
\end{equation}
Fig.~\ref{fig:excitation_config} displays in a schematic way the most important neutron 2-qp contributions to the isoscalar monopole excitation at $\hbar \omega = 6.7$ MeV. The 
single-particle levels correspond to the diagonal matrix elements of the single-particle Hamiltonian in the canonical basis, and the occupation numbers are the eigenvalues of the density matrix. We have
obtained almost identical results for the proton contributions. Obviously this excitation is only very weakly collective with just a few relevant 2-qp contributions. Among them, by far most significant is the transition from the almost fully occupied $1/2^+$ state that originates from the spherical $1d_{5/2}$ shell, to the unoccupied $1/2^+$ state based on the
spherical $2s_{1/2}$ shell. 
Such a 2-qp excitation can be considered in the context of spontaneous breaking of rotational symmetry which captures in an economic way non-trivial correlations as the source of collective behavior of the nucleus. This spontaneous breaking of rotational symmetry leads to the appearance of new excitation modes commonly referred to as a density wave \cite{kan11}. Density waves are related to the variation of the modulus of the order parameter of the broken symmetry.

\begin{figure}[!htb]
\centering
\includegraphics[width=0.75\linewidth]{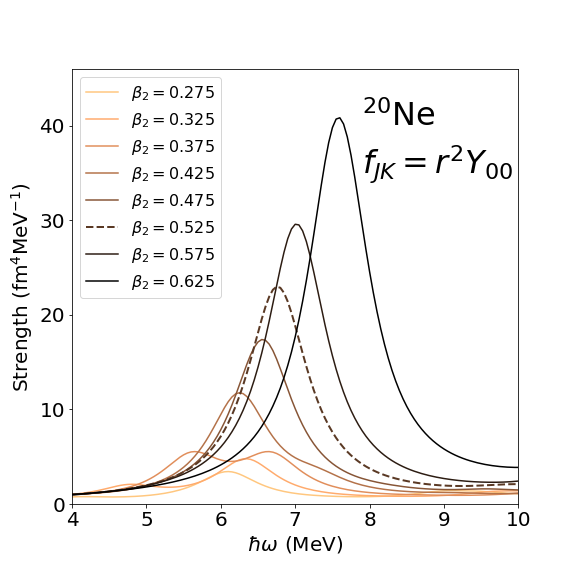}
\caption{(Color online) The low-energy isoscalar monopole strength distribution in $^{20}$Ne isotope. 
The QFAM response is calculated for several constrained values of the axial quadrupole deformation $\beta_2$, and
the dashed curve corresponds to the equilibrium deformation $\beta_2 = 0.525$. }
\label{fig:Strength_b2_20_0_0}
\end{figure}

Large deformations favor the formation of clusters \cite{ich11,yao14} and the previous discussion also suggests that there is a close link between 
cluster vibrational modes and nuclear deformation. The evolution of the low-energy cluster modes with deformation can be studied in more detail by performing a deformation-constrained calculation. In Fig.~\ref{fig:Strength_b2_20_0_0} we display the isoscalar monopole strength in $^{20}$Ne for several values of the axial quadrupole constraint, 
from $\beta_2 = 0.275$ to $\beta_2=0.625$. The dashed curve ($\beta_2=0.525$) corresponds to the strength distribution built on top of the mean-field equilibrium deformation. 
Significant strength in the region $\hbar \omega \approx 5-7$ MeV begins to appear at $\beta_2 \sim 0.2$ and, with increasing deformation, the fragmented strength evolves towards a single peak at slightly higher energy.

\begin{figure}[!htb]
\centering
\includegraphics[width=1\linewidth]{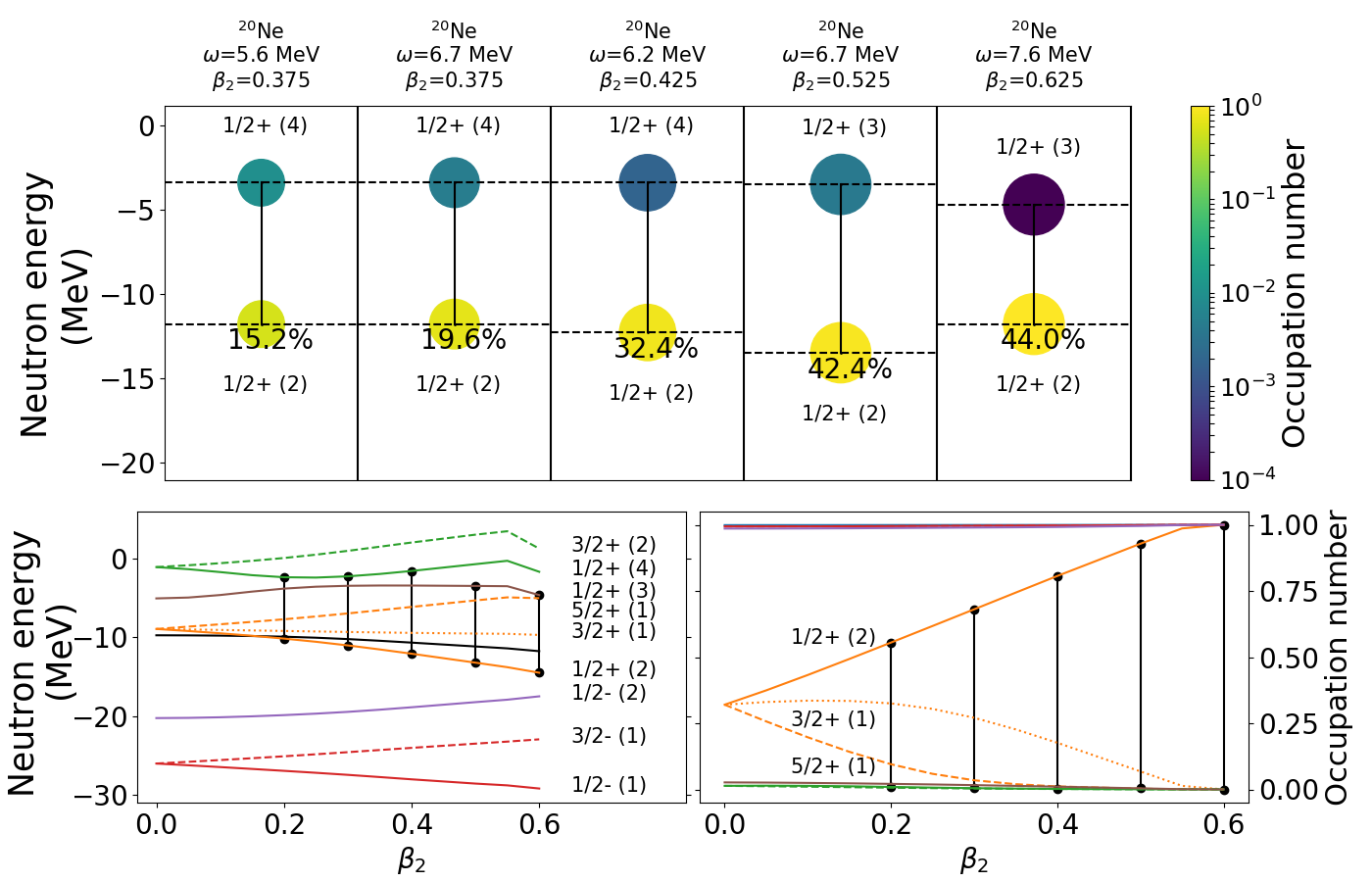}
\caption{(Color online) Evolution of the leading neutron 2qp contributions to the low-energy monopole mode with constrained deformation (upper panel). The lower panel shows the evolution of the single-particle energies (left) and occupation number (right) in the canonical basis with deformation. The vertical black lines denote the transitions that correspond to the principal 2-qp contribution shown in the upper panel. The thick black curve denotes the Fermi level.
}
\label{fig:XY_20Ne_def}
\end{figure}

\begin{figure}[!htb]
\centering
\includegraphics[width=1\linewidth]{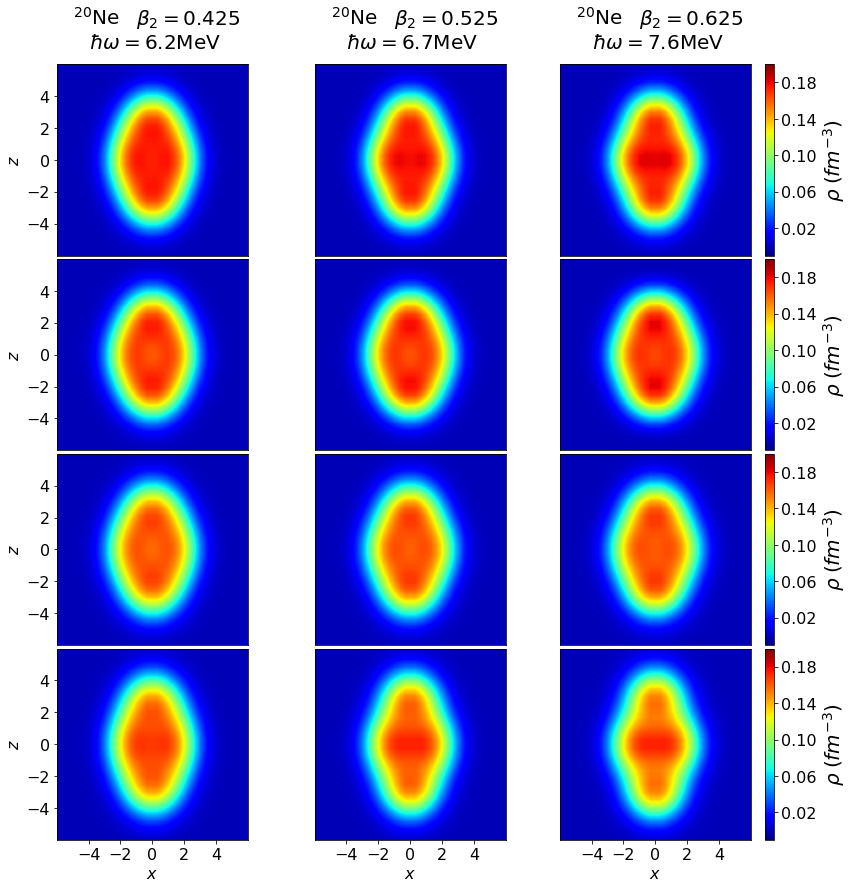}
\caption{(Color online) Snapshots of $^{20}$Ne total density monopole oscillations at energy $\hbar \omega$ and constrained initial deformation $\beta _2$. The time flows from the top to the bottom and a full period is shown.
}
\label{fig:dens_20Ne_J0_all}
\end{figure}

The appearance of cluster oscillations can be related to the structure of single-nucleon levels in the canonical basis. In the upper panel of Fig.~\ref{fig:XY_20Ne_def} we display the two largest neutron 2-qp contributions to the low-lying cluster vibration mode at the energy corresponding to a given constrained deformation (see also caption to Fig.~\ref{fig:excitation_config}). The lower panels show the evolution of the single-particle energies and occupation probabilities in the canonical basis. As the deformation increases the $1d_{5/2}$ spherical shell splits into three levels: $1/2^+$, $3/2^+$ and $5/2^+$. In particular, the occupation probability for the $1/2^+$ level increases with deformation thus enabling hole-particle excitations to the $1/2^+$ states originating from the spherical $2s_{1/2}$ and $1d_{3/2}$ shells. We note that the occupation of the $1/2^+$ level based on the $1d_{5/2}$ spherical shell is, of course,  also responsible for the formation of clusters in the ground state of $^{20}$Ne.
As shown in Fig.~\ref{fig:XY_20Ne_def}, the lowest deformation for which the low-energy monopole excitation is obtained is $\beta \approx 0.2$, which  coincides with the intersection of the $1/2^{+}[2 0 0]$ level and the Fermi level. A further increase of deformation between $\beta _2 =0.4$ and $\beta _2 = 0.5$  leads to a rearrangement of the contribution of the levels $1/2^+[0 1 0]$ and  $1/2^+[1 0 1]$ to the QFAM transition strength. The contribution of these levels to the total strength increases from 25\% to more than 40\%. The oscillations with constrained deformation are illustrated in Fig.\ref{fig:dens_20Ne_J0_all}, where we display the snapshots of the total density oscillations at energy $\hbar \omega$ and constrained deformation $\beta _2$ caused by a monopole  perturbation. At larger deformations the cluster structure is, of course, more pronounced. The oscillation frequency increases because the energy splitting of the single-particle levels increases with deformation.  

The very low-energy excitation at $\hbar \omega \approx 2$ MeV (se Fig.~\ref{fig:Ne20_strength_all}) can also be understood from the 1d5/2 splitting. It turns out that this excitation can be attributed to a  pure pairing effect due to the partial filling of the $1/2^+[2 0 0]$ and $3/2^+[1 0 1]$ levels. They are competing between $\beta _2 = 0$ and $\beta _2 = 0.5$, at which deformation the 1/2+[200] becomes fully occupied. Between these deformations, and because these levels are very close to the Fermi energy, pairing excitations can occur, depending on the pairing gap as well as the quasiparticle energies. 

%
%
\section{Isoscalar monopole response of $N=Z$ nuclei}
%

\begin{figure}[!htb]
\centering
\includegraphics[width=0.95\linewidth]{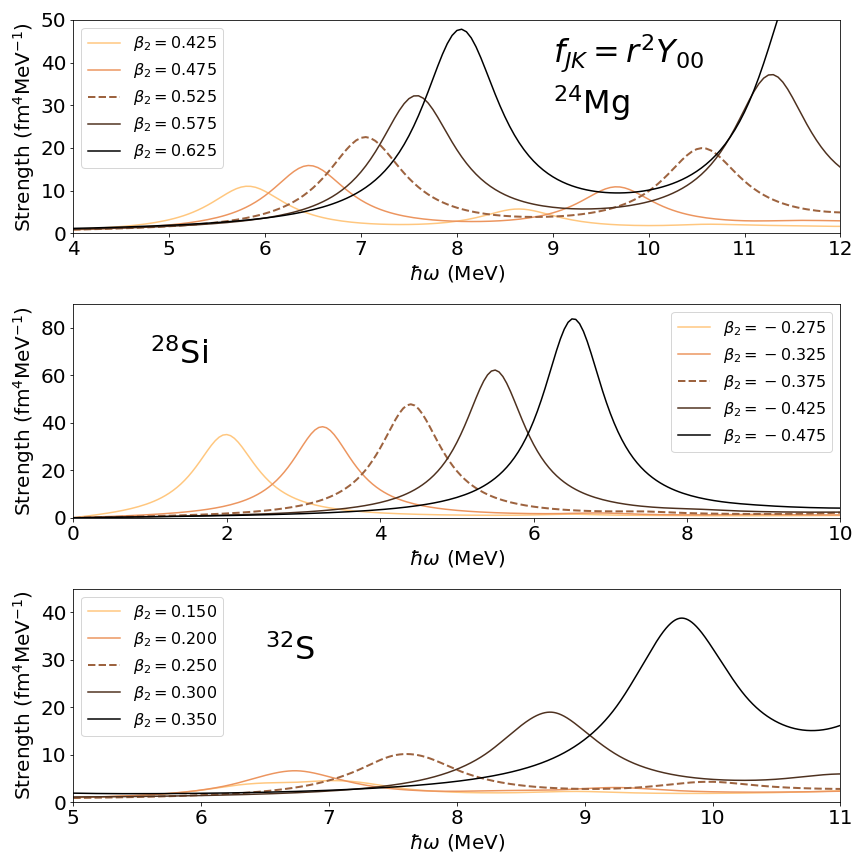}
\caption{(Color online) Low-energy isoscalar monopole strength distribution in $N=Z$ nuclei: $^{24}$Mg, $^{28}$Si and $^{32}$S.
The QFAM response is calculated for several values of constrained  axial quadrupole deformation $\beta_2$, and
the dashed curves correspond to the equilibrium deformation for each nucleus.}
\label{fig:monopole_strength-n_z}
\end{figure}

In this section we extend the analysis of low-lying isoscalar monopole QFAM response to 
 $^{24}$Mg, $^{28}$Si and $^{32}$S.
Figure~\ref{fig:monopole_strength-n_z} displays the corresponding isoscalar monopole strength functions for several values of the axial quadrupole constraint $\beta_2$.
One notices the appearance of the low-energy and large prolate deformation peak of the strength distribution for all isotopes shown in Fig.~\ref{fig:monopole_strength-n_z}, similar to the results obtained for $^{20}$Ne in the previous section. We have also performed corresponding calculations for other light and medium-heavy $N=Z$ nuclei, from $^{12}$C to $^{56}$Ni. 
The appearance of low-energy strength is much less pronounced for isotopes in the vicinity of doubly closed-shells.

\begin{figure}[!htb]
\centering
\includegraphics[width=1\linewidth]{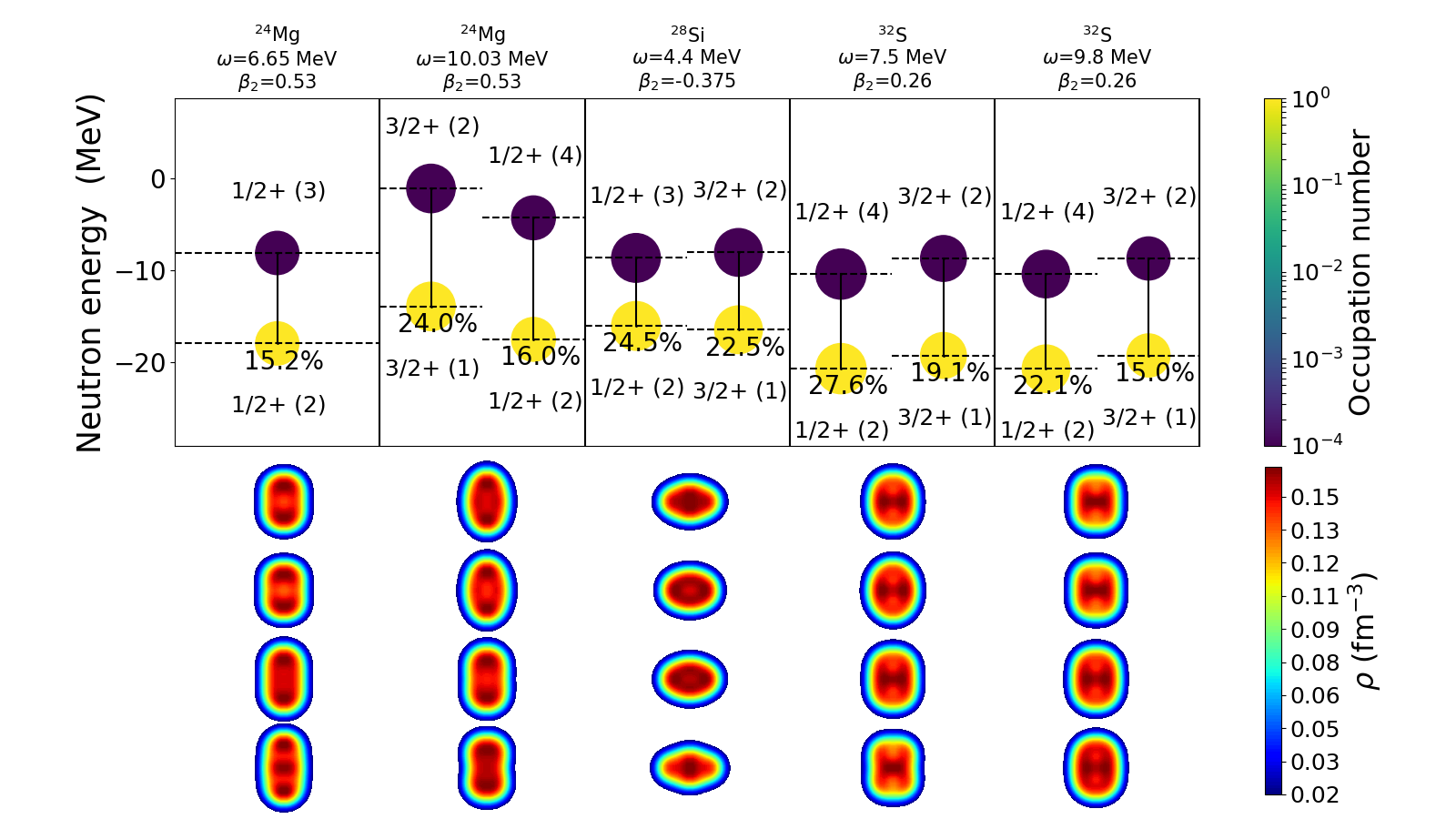}
\caption{ (Color online) Upper panel: leading neutron 2-qp contributions to the low-energy monopole modes in $^{24}$Mg, $^{28}$Si and $^{32}$S isotopes
 (for detailed description see the caption to Fig.~\ref{fig:XY_20Ne_def}). Lower panel:  snapshots of the corresponding density oscillations (see the caption to Fig.~\ref{fig:dens_20Ne_J0_all}). 
}
\label{fig:XY_NZ}
\end{figure}

The structure of the strength distributions can be analyzed by considering the principal 2-qp contributions, displayed in Fig.~\ref{fig:XY_NZ}.
We have selected several low-energy peaks in $^{24}$Mg, $^{28}$Si and $^{32}$S, and the results again indicate that these low energy excitations are primarily determined by a single 2-qp excitation.
In $^{24}$Mg we obtain two peaks, one at $\sim 7$ MeV and a second one at $\sim 10$ MeV, that have already been observed in experiment \cite{gup16}. Similar to the case of $^{20}$Ne, the lower state in $^{24}$Mg (first column of Fig.~\ref{fig:XY_NZ}) is mainly determined by the transition between the $1/2^+$ states originating from the $1d_{5/2}$ spherical shell (hole-like) and $2s_{1/2}$ spherical shell (particle-like). The addition of two neutron and two protons leads to the appearance of the second mode at excitation energy 
$\hbar \omega =10.03$ MeV (second column of Fig.~\ref{fig:XY_NZ}). This excitation, corresponding to the oscillations of two large clusters ($^{12}$C + $^{12}$C), is determined by the
transition between the $3/2^+$ states originating from the  $1d_{5/2}$ spherical shell (hole-like) and $1d_{3/2}$ spherical shell (particle-like). While for $^{20}$Ne the $3/2^+[101]$ state was not occupied, two more particles in $^{24}$Mg start filling the $3/2^+[101]$ state with the occupation probability approaching 1 for $\beta _2 \approx 0.7$. Hence, the mechanism that drives the low-energy excitations in $^{24}$Mg isotope is generally the same as for  $^{20}$Ne. The splitting of the spherical $1d_{5/2}$ and $1d_{3/2}$  levels with deformation allows now for two transitions, one between $\Omega^\pi=1/2^+$ states, and another between $\Omega^\pi=3/2^+$ states. 
Similar arguments apply to other low-energy excitations shown in Fig.~\ref{fig:XY_NZ}. 
%

%
%
%
\section{Summary and conclusion}
%
A systematic analysis of low-lying multipole response in deformed $N=Z$ nuclei has been performed using the quasiparticle finite amplitude method based on relativistic nuclear energy density functionals. It has been shown that the low-energy modes correspond to cluster vibrations for all considered isoscalar multipole operators. In particular, in  $^{20}$Ne the monopole and quadrupole operators induce oscillations of two $\alpha$-clusters around the $^{12}$C core, while the dipole and octupole operators induce vibrations of an $\alpha$-cluster with respect to the $^{16}$O core.

To analyze the effect of deformation on the low-lying strength distribution, in a first step we have performed a deformation-constrained QFAM calculation for the monopole response in $^{20}$Ne.  The appearance of cluster oscillations is closely related to the structure of single-nucleon levels in the canonical basis and, in particular, to the splitting of the $1d_{5/2}$ spherical shell. The monopole response is governed predominantly by the transition from the $1/2^+$ state originating from the spherical $1d_{5/2}$ shell to the $1/2^+$ state that correspond to the spherical $2s_{1/2}$ shell. 
We have also extended the analysis of the low-lying isoscalar monopole QFAM response for light and medium-heavy $N=Z$ nuclei,
from $^{12}$C to $^{56}$Ni. It has been found that the low-energy peaks of the monopole strength distribution are more pronounced in deformed isotopes far from closed shells.
The results are illustrated by three isotopes with clearly visible cluster vibration low-energy modes: $^{24}$Mg, $^{28}$Si and $^{32}$S. Similar to the $^{20}$Ne case, the
low-energy excitations in these isotopes are dominated by single 2-qp excitations. 
A study of higher-multipole QFAM response in light and medium-heavy $N=Z$ nuclei is in preparation. 

\acknowledgments
This work has been supported  in part by the QuantiXLie Centre of Excellence, a project co-financed by the Croatian Government and European Union through the European Regional Development Fund - the Competitiveness and Cohesion Operational Programme (KK.01.1.1.01).

\end{document}